\documentclass[preprint]{revtex4}

\usepackage{graphicx}
\newcommand{\sgn}{\,{\rm sgn}\,}
\newcommand{\mod}{\,{\rm mod}\,}
\newcommand{\Tr}[1]{\,\mathop{\rm Tr}_{#1}\,}
\renewcommand{\vec}[1]{\mbox{\boldmath $#1$}}
\begin{document}

\title{The path-integral analysis of an associative memory model storing an infinite number of finite limit cycles}
\author{Kazushi Mimura}
\affiliation{
Faculty of Information Sciences, Hiroshima City University, 
Ohtsukahigashi 3-4-1, Asaminami-ku, Hiroshima, 731-3194 Japan, \\
Kobe City College of Technology, 
Gakuenhigashi-machi 8-3, Nishi-ku, Kobe, Hyogo, 651-2194 Japan}
\email{mimura@cs.hiroshima-cu.ac.jp}
\author{Masaki Kawamura}
\affiliation{Faculty of Science, Yamaguchi University, Yoshida 1677-1,
Yamaguchi, 753-8512 Japan}
\author{Masato Okada}
\affiliation{
Laboratory for Mathematical Neuroscience,
RIKEN Brain Science Institute, Saitama 351-0198, Japan, \\
"Intelligent Cooperation and Control", PRESTO, JST, 
c/o RIKEN BSI, Saitama 351-0198, Japan, 
%ERATO Kawato Dynamic Brain Project, 
%2-2 Hikaridai, Seika-cho, Soraku-gun, Kyoto 619-0288, Japan 
}

\date{\today}

\begin{abstract}
It is shown that an exact solution of the transient dynamics of an associative memory model 
storing an infinite number of limit cycles with $l$ finite steps by means of the path-integral analysis. 
%有限ステップ系列想起モデルの想起過程を経路積分によって厳密に解析する．
%
Assuming the Maxwell construction ansatz, we have succeeded in deriving the stationary state equations of the order parameters 
from the macroscopic recursive equations with respect to the finite-step sequence processing model which has retarded self-interactions. 
%---modified---We derived the stationary state equations given by the path-integral analysis. 
%想起仮定の動特性から定常状態を記述する巨視状態方程式を想起過程から導出する．
%
We have also derived the stationary state equations by means of the signal-to-noise analysis (SCSNA). %---ADDED---
The signal-to-noise analysis must assume that crosstalk noise of an input to spins obeys a Gaussian distribution. 
On the other hand, the path-integral method does not require such a Gaussian approximation of crosstalk noise. 
We have found that both the signal-to-noise analysis and the path-integral analysis 
give the completely same result with respect to the stationary state in the case where the dynamics is deterministic, when we assume the Maxwell construction ansatz. %---ADDED---
%この定常状態を記述する巨視状態方程式は，以前我々が求めたS/N解析による解析結果と一致した．
%このように，S/N解析ではクロストークノイズがガウス分布に従うと仮定するにもかかわらず，
%定常状態においてはガウス近似が不必要な経路積分による解析結果と一致した．
%
%---DELETED---This means that the result of the signal-to-noise analysis (SCSNA) is an exact solution in the case where $T=0$. 
%つまり，S/N解析による定常状態の解析結果が厳密解であることがわかった．
%
\par
We have shown the dependence of storage capacity ($\alpha_c$) on the number of patterns per one limit cycle ($l$). 
%また，定常状態を記述する巨視状態方程式から，系列のステップ数と記憶容量の関係を求めた．
%
At $l=1$, storage capacity is $\alpha_c=0.138$ like the Hopfield model's. 
%ステップ数が$l=1$のときは，記憶容量はHopfieldモデルと同じ$\alpha_c=0.138$となる．
%
Storage capacity monotonously increases with the number of steps, and converges to $\alpha_c=0.269$ at $l\simeq 10$. 
%ステップ数の増加とともに単調に記憶容量は増加し，
%ステップ数$l=10$程度で，すぐに無限ステップ系列想起モデルの記憶容量$\alpha_c=0.269$に収束する．
%
The original properties of the finite-step sequence processing model appear as long as the number of steps of the limit cycle has order $l=O(1)$. 
%ステップ数$l=O(1)$でのみ有限ステップに特有の性質を持つことがわかった．
\end{abstract}

\maketitle

\section{Introduction}
%~~~~~~~~~~~~~~~~~~~~~~~~~~~~~~~~~~~~~~~~~~~~~~~~~~~~~~~~~~~~~~~~~~~~~

%無限系列想起モデル　infinite-step sequence processing model
%有限系列想起モデル　  finite-step sequence processing model

In recent years, theories that can analyze the transient dynamics have been discussed for systems with frustrations especially a correlation-type associative memory 
\cite{hopfield1982,shiino1992,amit1985b,kuhn1991,mimura1995,okada1996,mimura1996,mimura1998a,mimura1998b,mimura2003,okada1995}. 
%近年，フラストレーションを持つ系，特に相関型連想記憶モデルに関して動的理論の議論が盛んに行われている\cite{hopfield1982}-\cite{okada1995}．
%
D${\rm \ddot{u}}$ring et al presented a path-integral method for an infinite-step sequence processing model and analyzed the properties of the stationary state \cite{during1998}. 
%Duringは無限ステップ系列想起モデルに経路積分を適用し，その定常状態方程式から記憶容量などを求めた\cite{during1998}．
%
By using D${\rm \ddot{u}}$ring et al's analysis, Kawamura and Okada succeeded in deriving an exact macroscopic description of the transient dynamics \cite{kawamura2002}. 
%KawamuraとOkadaはDuringの定式化に基づき過渡特性を記述する巨視状態方程式を求めた\cite{kawamura2002}．
%
The transient dynamics can be analyzed not only by using the path-integral method \cite{coolen1993,coolen2000} but by using the signal-to-noise analysis, e.g., statistical neurodynamics \cite{amari1988a,amari1988b}．
%連想記憶モデルの動特性は，経路積分\cite{coolen1993,coolen2000}だけでなく統計神経力学と呼ばれるS/N解析によっても解析できる\cite{amari1988a,amari1988b}．
%
The signal-to-noise analysis is an approximation theory in which crosstalk noise obeys a Gaussian distribution. 
%統計神経力学は，クロストークノイズをガウス分布で近似して巨視状態を表す漸化式を導く近似理論である．
%
On the other hand, the path-integral method does not require such a Gaussian approximation of crosstalk noise. 
%一方，経路積分による解析では，クロストークノイズのガウス近似は不必要である．
%
However surprisingly, the macroscopic equations of the exact solution given by means of the path-integral method are completely equivalent to those of the signal-to-noise analysis with respect to this model. 
%にもかかわらず，驚くべきことに無限ステップ系列想起モデルについては，
%厳密解である経路積分による結果と，近似理論であるS/N解析の結果は完全に一致する．
%
\par
It has turned out that the infinite-step sequence processing model can be more easily analyzed than the Hopfield model even if it is necessary to treat the dynamical process directly. 
%無限ステップ系列想起モデルでは，動的な過程を直接取扱う必要があるにも関わらず，自己想起モデルよりも理論的な取扱いが容易である．
%
This reason is as follows. 
The retrieval state of the infinite-step sequence processing model has no equilibrium state. 
Therefore, the correlations of the system are not very complex. 
%これは，系列想起モデルでは同じ記憶パターンを回路の状態として2度以上取らないために，系の統計的な性質が簡単になるためである．
%
Since the Hopfield model takes the same states repeatedly, 
its statistical properties are more complex than the infinite-step sequence processing model. 
%しかし，自己想起モデルでは，系が同じ状態を何度もとるために，時間的な相関が複雑になる．
%
Gardner et al analyzed the transient dynamics of the Hopfield model by using the path-integral method in the case where the dynamics is deterministic \cite{gardner1987}. 
%Gardnerらは，経路積分的な方法により，自己想起モデルのダイナミクスを解析した．
%
They obtained the macroscopic equations of the transient dynamics at time step $t$ using $O(t^2)$ macroscopic variables and also obtained the macroscopic equations of the equilibrium state from the transient dynamics. 
%決定論的ダイナミクスの場合について，Gardnerらの解析は，任意の$t$ステップでの$O(t^2)$個の巨視的変数を使って巨視的状態方程式を得た．
%
These are equivalent to replica symmetric (RS) solutions given by using the replica method \cite{gardner1987}. 
%ガードナーらの結果は，レプリカ法のRS解に一致する\cite{gardner1987}．
%
%---DELETED--- On the other hand, it is well known that the RS solutions are equivalent to those of the signal-to-noise analysis \cite{shiino1992}. 
%---DELETED--- %一方，RS解がs/n解析による解と一致することがよく知られている．
%---DELETED--- %
%---DELETED--- However until now, the relation between the path-integral method and the signal-to-noise analysis was unknown, 
%---DELETED--- because of the complexity of the statistical property. 
%---DELETED--- %これまで，統計的性質が複雑で取り扱いの難しい（自己想起モデルについては），経路積分とS/N解析の関係は不明であった．
%---DELETED--- %
Recently in the Hopfield model, Bolle et al compared the transient dynamics of the path-integral method with those of the signal-to-noise analysis only for a few time steps in the dynamics \cite{bolle2003}. 
%ごく最近，Bolleらは，自己想起モデルをS/N解析と経路積分で解析して結果を比較した\cite{bolle2003}．
%
They have pointed out that the signal-to-noise analysis is exact up to time step 3 and inexact to step 4 or beyond. 
%彼らは，想起過程の3ステップ目までは両者の結果は一致し，4ステップ目以降はずれが生じるとの結果を報告している．
%
\par
In order to discuss the relation between the path-integral method and the signal-to-noise analysis in more detail, 
we analyze a finite-step sequence processing model, which includes the Hopfield model and the infinite-step sequence processing model in special cases. 
%そこで本論文では，無限ステップ系列想起モデルと自己想起モデルを特殊な場合として含む，
%有限ステップ系列想起モデルを経路積分よって解析し，自己想起モデルでの経路積分とS/N解析の関係を議論する．
%
In the finite-step sequence processing model, the steady states of the system become limit cycles. 
%有限ステップ系列想起モデルでは，系の定常状態はリミットサイクルとなる．
%
Since the finite-step sequence processing model can store limit cycles in the dynamics, 
the properties of the system are periodic and dynamic essentially like the infinite-step sequence processing model. 
%有限ステップ系列想起モデルはリミットサイクルを記憶させているために，
%無限ステップ系列想起モデルと同様に系の性質は本質的に動的である．
%
Moreover, the statistical properties of the finite-step sequence processing model are more complex than the infinite-step one. 
Since the period of the limit cycle is finite, the network takes the same states repeatedly. 
%またリミットサイクルの周期が有限であるために，回路の同じ状態を何度も取るため，自己想起モデルと同様の統計的相関が生じる．
%
Namely, the finite-step sequence processing model has the theoretical difficulties of both the Hopfield model and the infinite-step sequence processing model. 
%つまり，有限ステップ系列想起モデルは，自己想起モデルと系列想起モデルの理論的な難しさを両方持っている．
%
In this point of view, it would be very interesting to theoretically discuss the properties of the finite-step sequence processing model. 
%この観点においても，このモデルの性質を理論的に議論することは大変興味深い．
%
\par
In this paper, we have exactly derived the transient dynamics of macroscopic recursive equations 
with respect to the finite-step sequence processing model by means of the path-integral analysis. %---ADDED---
%経路積分法によって得られる結果は，巨視状態のtransient dynamics である．
%
Until now, only in the infinite-step sequence processing model, which has no self-interactions, 
D${\rm \ddot{u}}$ring et al derived the stationary state equations of the order parameters by using the path-integral analysis \cite{during1998}. %---ADDED---
%これまで，extensive loading の場合($\alpha \ne 0$)については，transient dynamics から，
%定常状態方程式を導出は infinite-step sequence processing model においてしか成功していない \cite{during1998}．
%
The transient dynamics of various disordered systems can be also analyzed by using the path-integral method. 
%In the recent research, the path-integral analysis is adopted to various problems, e.g., minority game \cite{coolen2001}, on-line learning \cite{inoue2001} and so on. 
%経路積分法は，非常に一般的な解析方法であり，最近ではマイノリティゲーム\cite{coolen2001}やonline学習理論\bibitem{inoue2001}などへの適用もすでになされている．
%
Therefore, it is important to derive stationary state equations of the order parameters from the macroscopic recursive equations. 
%この観点からも，Maxwell construction を利用するとはいうものの，transient dynamics からの定常状態の導出は重要である．
%
Assuming the Maxwell construction ansatz, we have succeeded in deriving the stationary state equations 
from the macroscopic recursive equations with respect to the model, which has self-interactions, i.e., the finite-step sequence processing model. %---ADDED---
%我々は retarded self-interaction がある場合についても定常状態方程式を導いた．
%
%---DELETED---In the case where the dynamics is deterministic, we also derived the macroscopic equations for stationary states. 
\par
We also analyzed the finite-step sequence processing model by means of the signal-to-noise analysis (SCSNA). %\cite{mimura1995}
The stationary state equations given by the path-integral analysis are equivalent to those of the signal-to-noise analysis. 
%有限ステップ系列想起モデルを経路積分によって解析して得られた結果は，
%$t\to\infty$,$T\to0$では，以前我々が求めたS/N解析による解析結果と一致する．
%
%---DELETED---It is very interesting that the path-integral method, which does not require a Gaussian approximation of crosstalk noise, 
%---DELETED---and the signal-to-noise analysis, which needs such an approximation, give the same results in the stationary state. 
%S/N解析では，クロストークノイズのガウス近似が必要であるにも関わらず，
%定常状態においてはガウス近似が不必要な経路積分による解析結果と一致することはとても興味深い．
%
This result corresponds to the fact that the replica method and the signal-to-noise analysis give completely equivalent results in the stationary state analysis of the Hopfield model. 
%これは，これまでの自己想起モデルの定常状態の解析で，
%レプリカ法などによる解析結果と，S/N解析による解析結果が一致することに対応している．
%
Namely, the transient dynamics given by the signal-to-noise analysis gives an exact solution in both the stationary state and the first few time steps in the dynamics. 
%つまり，統計神経力学などのS/N解析による動特性の解析により，
%初期の数ステップと定常状態に関しては経路積分と同様に厳密に解析できることがわかった．
%
%---DELETED--- This includes Gardner et al's result concerning the Hopfield model, i.e., 
%---DELETED--- the property of the equilibrium state given by a similar way to the path-integral method 
%---DELETED--- is equivalent to those of the signal-to-noise analysis \cite{gardner1987}. 
%---DELETED--- %この結果は，Gardnerの結果を含んでいる．
%---DELETED--- %

\section{Definitions}
%~~~~~~~~~~~~~~~~~~~~~~~~~~~~~~~~~~~~~~~~~~~~~~~~~~~~~~~~~~~~~~~~~~~~~

Let us consider a system storing an infinite number of limit cycles with $l$ finite steps.
The system consists of $N$ Ising-type spins (or neurons) $\sigma_i=\pm 1$. 
We consider the case where $N \to \infty$. 
The spins updates the state synchronously with the probability: 
\begin{equation}%(dynamics)
{\rm Prob}[\sigma_i(t+1)=-\sigma_i(t)]=\frac 12 \biggl[ 1 - \sigma_i(t) \tanh \beta h_i(t) \biggr] ,
\label{eq:dynamics}
\end{equation}
\begin{equation}%(internal potential)
h_i(t)=\sum_{j=1}^N J_{ij} \sigma_j(t) + \theta_i(t) ,
\end{equation}
where $\beta$ is the inverse temperature, $\beta = 1/T$. 
When the temperature is $T=0$, the updating rule of the state is deterministic. 
The term $\theta_i(t)$ is a time-dependent external field which is introduced in order to define a response function. 
The interaction $J_{ij}$ stores $p$ random patterns $\vec{\xi}^{\nu,\mu}=(\xi_1^{\nu,\mu},\cdots,\xi_N^{\nu,\mu})^T$ so as to retrieve the patterns as: 
\begin{equation}
\vec{\xi}^{\nu,1} \to \vec{\xi}^{\nu,2} \to \cdots \to \vec{\xi}^{\nu,l} \to \vec{\xi}^{\nu,1}, \nonumber
\end{equation}
sequentially for any $\mu$th limit cycle. 
For instance, the entries of the interaction matrix $\vec{J}=(J_{ij})$ are given by 
\begin{equation}%(the entries of the interaction matrix)
J_{ij}=\frac 1N \sum_{\nu =1}^{p/l} \sum_{\mu =1}^l \xi_i^{\nu ,\mu +1} \xi_j^{\nu ,\mu},
\end{equation}
where the pattern index $\mu$ are understood to be taken modulo $l$. 
Since the number of limit cycles is $p/l$, the total number of stored patterns is $p$. 
The number of stored patterns $p$ is given by $p=\alpha N$, where $\alpha$ is called the loading rate. 
In our analysis, the number of steps for each limit cycle $l$ is kept finite. 
Each component of the patterns is assumed to be an independent random variable that takes a value of either $+1$ or $-1$ according to the probability: 
\begin{equation}%(each component of the patterns)
{\rm Prob}[\xi_i^{\nu,\mu}=\pm 1]=\frac 12.
\end{equation}
%We determine the initial state $\vec{\sigma}(0)$ according to the following probability distribution 
%\begin{equation}%(initial state)
%{\rm Prob}[\sigma_i(0)=\pm 1]=\frac {1\pm m(0)\xi_i^{1,l}}2.
%\end{equation}
%Therefore, the overlap between the pattern $\vec{\xi}^{1,l}$ and the initial state $\vec{\sigma}(0)$ is $m(0)$. 
%The matrix $\vec{J}$ is not non-symmetric. 
For the subsequent analysis, the matrix $\vec{J}$ is represented as 
\begin{equation}%(interaction matrix)
\vec{J} = \frac 1N \vec{\xi}^T \vec{S} \vec{\xi},
\end{equation}
where the $p \times N$ matrix $\vec{\xi}$ is defined as 
\begin{equation}
\vec{\xi}=(\vec{\xi}^{1,1} \cdots \vec{\xi}^{1,l} \vec{\xi}^{2,1} \cdots \vec{\xi}^{2,l} \cdots \vec{\xi}^{p/l,1} \cdots \vec{\xi}^{p/l,l})^T , 
\end{equation}
and the $p \times p$ matrix $\vec{S}$ is defined as 
\begin{equation}%S
\vec{S}=
\left(
\begin{array}{ccc}
\vec{S}' & \cdots & 0        \\
         & \ddots &          \\
0        & \cdots & \vec{S}' \\
\end{array} 
\right) ,
\end{equation}
and the $l \times l$ matrix is defined as $\vec{S}'= (S_{\mu \nu}') = (\delta_{\mu, (\nu +1) \mod l})$. 
When $l=1$, i.e., $\vec{S}=\vec{1}$ ($\vec{1}$ is the unity matrix), the system is equivalent to the Hopfield model.

\section{Path-integral analysis}
%~~~~~~~~~~~~~~~~~~~~~~~~~~~~~~~~~~~~~~~~~~~~~~~~~~~~~~~~~~~~~~~~~~~~~

D${\rm \ddot{u}}$ring et al discussed the sequential associative memory model by means of the path-integral analysis \cite{during1998}. 
In this section, we introduce macroscopic state equations for the model with a finite temperature $T \ge 0$, according to their paper. 

In order to analyze the transient dynamics, the generating function $Z[\vec{\psi}]$ is defined as 
%まず，次の生成関数$Z[\vec{\psi}]$を定義する．
%
\begin{equation}%(definition of generating functional Z[phi])
Z[\vec{\psi}]=\sum_{\vec{\sigma}(0),\cdots ,\vec{\sigma}(t)} p[\vec{\sigma}(0),\cdots ,\vec{\sigma}(t)] e^{-i\sum_{s<t}\vec{\sigma}(s)\cdot \vec{\psi}(s)}, 
\label{eq:def_Z}
\end{equation}
where $\vec{\psi}=(\psi(0),\cdots ,\psi(t-1))^T$ and the state $\vec{\sigma}(s)=(\sigma_1(s),\cdots ,\sigma_N(s))^T$ denotes the state of the spins at time $s$. 
%ここで，$\vec{\psi}=(\psi(0),\cdots ,\psi(t-1))^T$であり，$\vec{\sigma}(s)=(\sigma_1(s),\cdots ,\sigma_N(s))^T$は時刻$s$でのネットワークの状態を示す．
%
The probability $p[\vec{\sigma}(0),\cdots ,\vec{\sigma}(t)]$ denotes the probability of taking the path from initial state $\vec{\sigma}(0)$ to state $\vec{\sigma}(t)$ at time $t$ through $\vec{\sigma}(1),\vec{\sigma}(2),\cdots,\vec{\sigma}(t-1)$. 
%また，$p[\vec{\sigma}(0),\cdots ,\vec{\sigma}(t)]$は初期状態$\vec{\sigma}(0)$から状態$\vec{\sigma}(1),\vec{\sigma}(2),\cdots$を経由して状態$\vec{\sigma}(t)$になる確率を示す．
%
As (\ref{eq:def_Z}) shows, the generating functional entails the summation of all $2^{(t+1)N}$ paths which the system can take from time $0$ to $t$. 
%(\ref{eq:def_Z})式からもわかるように，生成関数は時刻$0$から時刻$t$までにネットワークが取り得る$2^{(t+1)N}$個の経路に関する和をとっている．%このため，この生成関数による解析方法は経路積分法と呼ばれている．
%
One can obtain all the relevant order parameters, i.e., the overlap $m(s)$, the correlation function $C(s,s')$ and the response function $G(s,s')$, by calculating the appropriate derivatives of the above functional and letting $\vec{\psi}$ tend to $\vec{0}$ afterwards as follows: 
%生成関数$Z[\vec{\psi}]$よりモデルの巨視状態を表す，想起中のパターンと状態との重なりを表すオーバラップ$m(s)$，状態の時間相関$C(s,s')$，状態変化のしやすさを表す帯磁率$G(s,s')$を求めることができる．
%
\begin{equation}%(definition of m(s))
m(s)=i \lim_{\vec{\psi}\to\vec{0}} \frac 1N \sum_{i=1}^N \xi^{1,s}_i \frac{\partial Z[\vec{\psi}]}{\partial \psi_i(s)}, 
\label{eq:def_m}
\end{equation}
\begin{equation}%(definition of C(s,s'))
C(s,s')=- \lim_{\vec{\psi}\to\vec{0}} \frac 1N \sum_{i=1}^N \frac{\partial^2 Z[\vec{\psi}]}{\partial \psi_i(s) \partial \psi_i(s')},
\label{eq:def_C}
\end{equation}
\begin{equation}%(definition of G(s,s'))
G(s,s')=i \lim_{\vec{\psi}\to\vec{0}} \frac 1N \sum_{i=1}^N \frac{\partial^2 Z[\vec{\psi}]}{\partial \psi_i(s)\partial \theta_i(s')}. 
\label{eq:def_G}
\end{equation}
Using the assumption of self-averaging, we replace the generating functional $Z[\vec{\psi}]$ with its disorder-averaged generating functional $\bar{Z}[\vec{\psi}]$. 
Evaluating the averaged generating function $\bar{Z}[\vec{\psi}]$ through the saddle point method, 
we obtain the following saddle-point equations for the order parameters of (\ref{eq:def_m})-(\ref{eq:def_G}) in the thermodynamical limit, i.e., $N\to\infty$ (See Appendix \ref{app:-1}). 
\begin{eqnarray}
m(s)    &=& \ll \xi \sigma(s) \gg , \label{eq:spe_m} \\
C(s,s') &=& \ll \sigma(s) \sigma (s') \gg , \label{eq:spe_C} \\
G(s,s') &=& \frac{\partial \ll \sigma(s) \gg}{\partial \theta(s')}. \label{eq:spe_G}
\end{eqnarray}
The average over the effective path measure is given by 
\begin{equation}
\ll g(\vec{\sigma},\vec{v}) \gg 
= \biggl< \int {\cal D}\vec{v} \Tr{\vec{\sigma}} g(\vec{\sigma},\vec{v}) p[\sigma(0)] \prod_{s=1}^{t} \frac 12 [1+\sigma(s)\tanh \beta h (\vec{\sigma},\vec{v},s-1)] \biggr>_{\vec{\xi}}, 
\label{eq:E_path}
\end{equation}
\begin{eqnarray}
{\cal D}\vec{v} &\equiv& \frac {d\vec{v} e^{-\frac 12\vec{v} \cdot \vec{R}^{-1} \vec{v}}} {\sqrt{(2\pi )^t|\vec{R}|}}, \\ 
\Tr{\vec{\sigma}} &\equiv& \sum_{\sigma(0) ,\cdots ,\sigma(t) \in \{ -1,1 \}}, \\
h(\vec{\sigma},\vec{v},s) &=& \xi ^{s+1} m(s)+\theta(s)+\sqrt{\alpha}v(s)+(\vec{\Gamma} \vec{\sigma})(s), \label{eq:local_field} \\
\vec{R} &=& \sum_{a=0}^{l-1} \sum_{m,n \ge 0}\vec{G}^{ml+a} \vec{C}(\vec{G}^\dagger)^{nl+a}, \label{eq:R} \\
\vec{\Gamma} &=& \frac \alpha l \sum_{\mu =0}^{l-1} e^{2\pi i \mu / l} [\vec{1}-e^{2\pi i \mu / l}\vec{G}]^{-1}, \label{eq:gamma} 
\end{eqnarray}
with $\vec{\Gamma}=\hat{\vec{K}}^\dagger$, $\hat{\vec{Q}}=-\frac 12 \alpha i \hat{\vec{R}}$ 
and $p[\sigma(0)]=\frac 12[1+\sigma(0)m(0)]$ which is the initial spin probability. 
The operator $<\cdot>_{\vec{\xi}}$ denotes the average over the condensed patterns. 
The term $(\vec{\Gamma} \vec{\sigma})(s)$ denotes the $s$th element of the vector $\vec{\Gamma} \vec{\sigma}$. 
The vectors $\vec{\sigma}$ and $\vec{v}$ denote $\vec{\sigma}=\{ \sigma(0),\cdots , \sigma(t) \}$ and $\vec{v}=\{ v(0),\cdots , v(t-1) \}$, respectively. 
Equations (\ref{eq:spe_m})-(\ref{eq:gamma}) entirely describe the dynamics of the system. 
The term $\prod_{s=1}^{t} \frac 12[1+\sigma(s)\tanh \beta h (\vec{\sigma},\vec{v},s-1)]$ in (\ref{eq:E_path}) cannot be factorized with respect to spin variables at different times. 
Calculation of the spin summation of (\ref{eq:E_path}) requires an exponential time $O(e^t)$ at time $t$. 
In the infinite-step sequence processing model, local field $h(\vec{\sigma},\vec{v},s)$ depends on only spin variables at time $s$ \cite{during1998}. 
Therefore the term $\prod_{s=1}^{t} \frac 12 [1+\sigma(s)\tanh \beta h (\vec{\sigma},\vec{v},s-1)]$ can be factorized, so the spin summations can be taken easily.

\section{The stationary state}
%~~~~~~~~~~~~~~~~~~~~~~~~~~~~~~~~~~~~~~~~~~~~~~~~~~~~~~~~~~~~~~~~~~~~~

\begin{figure}[htbp]
\begin{center}
\includegraphics[width=.6\linewidth,keepaspectratio]{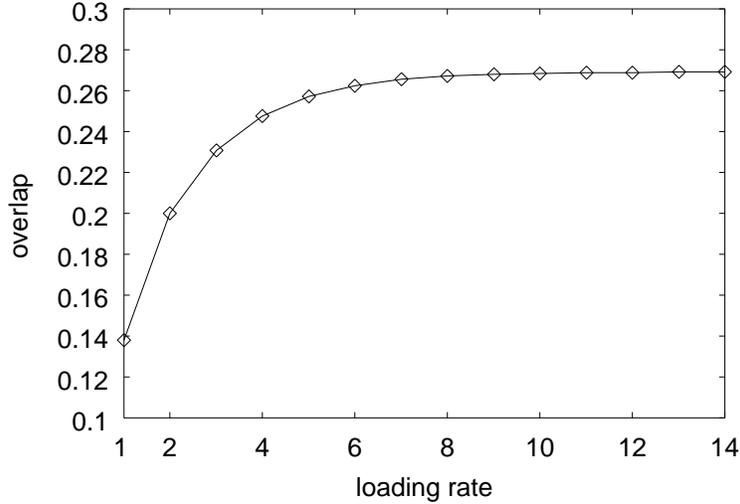}
\caption{the storage capacity $\alpha_c$ and the number of patterns per one limit cycle $s$.}
\label{fig:fig1}
\end{center}
\end{figure}

\begin{figure}[htbp]
\begin{center}
\includegraphics[width=.6\linewidth,keepaspectratio]{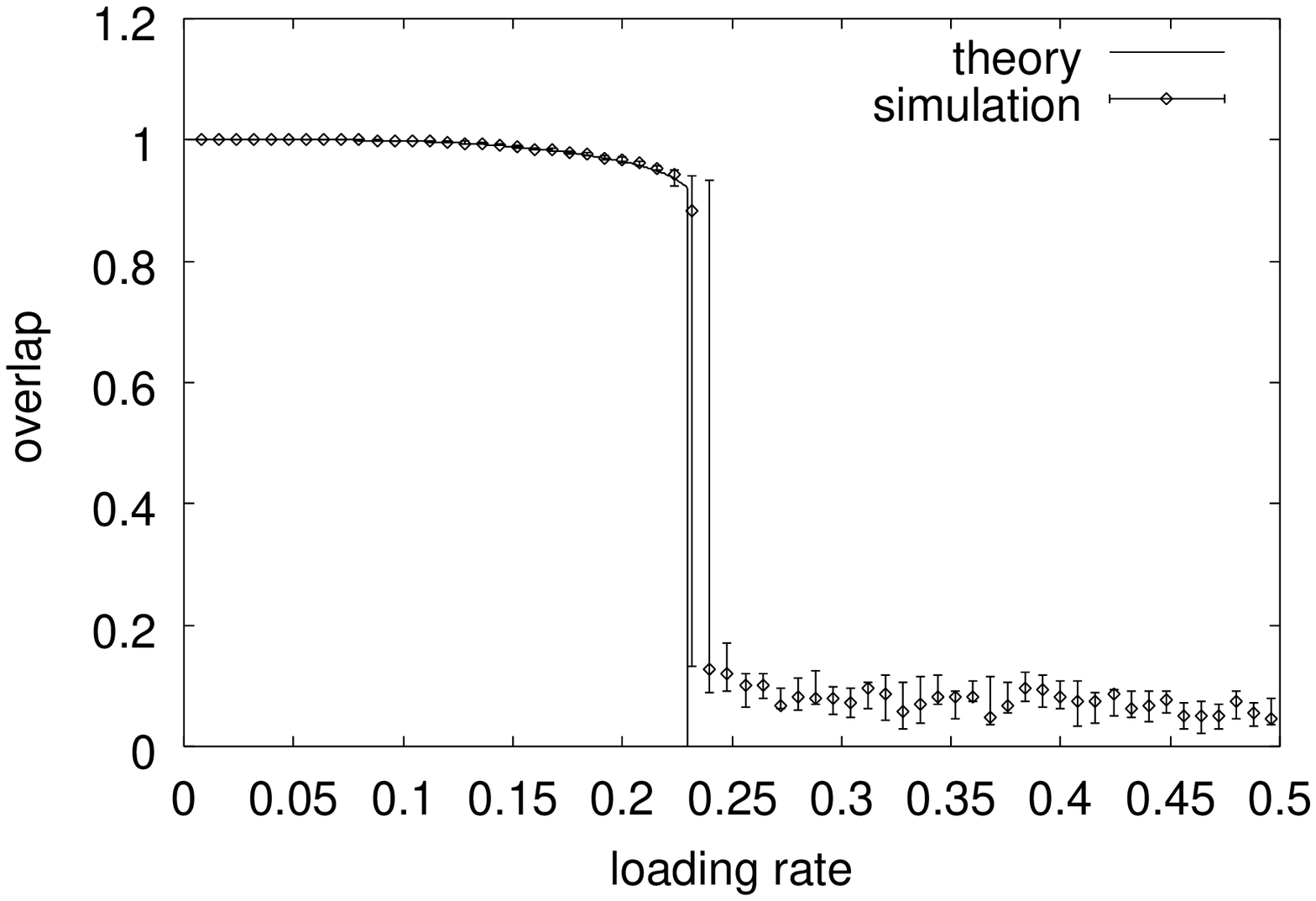}
\caption{Computer simulations ($l=3$,$N=3000$, $11$ times): the overlap $m$ and loading rate $\alpha$.}
\label{fig:fig2}
\end{center}
\end{figure}

\begin{figure}[htbp]
\begin{center}
\includegraphics[width=.6\linewidth,keepaspectratio]{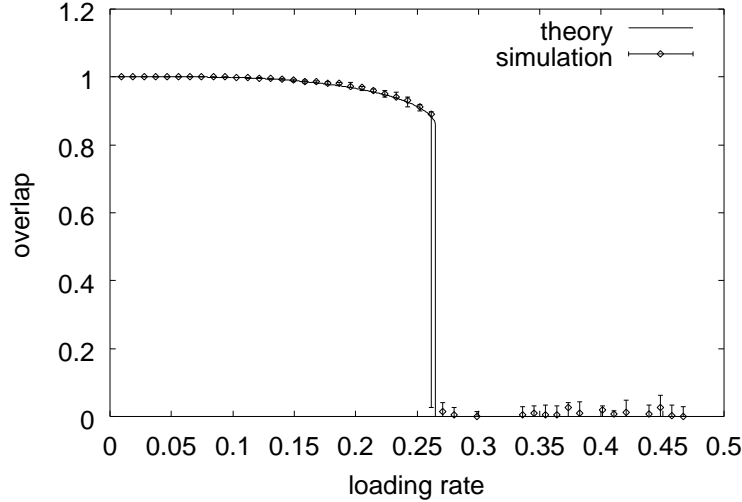}
\caption{Computer simulations ($l=7$, $N=3000$, $11$ times): the overlap $m$ and loading rate $\alpha$.}
\label{fig:fig3}
\end{center}
\end{figure}

In this section, we inspect time-translation invariant solutions of our macroscopic equations (\ref{eq:spe_m})-(\ref{eq:spe_G}) for the deterministic dynamics, i.e., $\beta\to\infty$ ($T=0$). 
The time-translation invariant solutions will describe motion on a macroscopic limit cycle: 
\begin{equation}
\left\{
\begin{array}{l}
m(s)=m, \\
C(s,s')=C(s-s'), \\
G(s,s')=G(s-s'), \\
\end{array}
\right.
\label{eq:condition}
\end{equation}
with $\theta(s)=\theta$. 
Now, we disregard the transient states. %---ADDED---
Note that the condition of (\ref{eq:condition}) includes an unspoken condition that the transient states are disregarded. %---ADDED---
Therefore, we put that the dynamics is already in the stationary state at time $s=0$ under this assumption. 
In the zero noise limit, i.e., $T=0$ ($\beta\to\infty$), the dynamics becomes deterministic. 
Therefore, we also assume that the system takes a fixed path as 
%---DELETED---In the stationary state, the system takes the same states repeatedly; therefore the dynamics of the system takes a fixed path. 
%つまり，定常状態である時刻$s \ge \tau$では，経路は1つに限定される．
%
%---DELETED---Namely, the time $\tau$, which satisfies following relationships, exists: 
%定常状態では，系の状態は周期$l$で同じ状態を繰り返しているため，...となる時刻$s=\tau$が存在する．
%
\begin{equation}
\sigma(s+l)=\sigma(s),
\end{equation}
for any time $s \ge 0$. 
The path which the system takes after time $s\ge 0$ can be described as 
\begin{equation}
\sigma(s)=\eta_s, \label{eq:stationary_state}
\end{equation}
by only $l$ constants $\eta_0,\cdots,\eta_{l-1}\in\{-1,1\}$. 
%$\sigma(\tau)=\sigma$とすると，$s\ge\tau$以降の経路は定数$\eta_0,\cdots,\eta_{l-1}\in\{-1,+1\}$によって，$\sigma(s+\tau')=\eta_{\tau'}\sigma$とおくことができる．
%
The pattern index $s$ of the constants $\eta_s$ is understood to be taken modulo $l$. 
%ここで，$\eta_s$の添え字は $\mod l$をとるとする．
%
Note that it is not necessary to calculate these constants $\{\eta_s\}$ explicitly. 
%この定数を求める必要はないことに注意されたい．
%
When the variable transformation 
\begin{equation}
\chi (s) = \eta_s \sigma (s), 
\label{eq:transformation}
\end{equation}
is carried out to spin variables $\sigma (s)$, 
the transformed spin variables $\chi(s)$ take same value for any time $s$, i.e., $\chi(s)=\chi(s')$ for any $s,s'$. 
Equation (\ref{eq:E_path}) means the expectation of $g(\vec{\sigma},\vec{v})$ with respect to the path probability. 
%(\ref{eq:E_path})式の中の項$\Tr{\vec{\sigma}}$は，$g(\vec{\sigma},\vec{v})$の path probability に関する期待値であり，
%
%---DELETED---In the zero noise limit, the dynamics defined by (\ref{eq:dynamics}) is equivalent to $\sigma_i(t+1)=\sgn h_i(t)$. 
%決定論的ダイナミクスのとき，(\ref{eq:dynamics})式のダイナミクスは，$\sigma_i(t+1)=\sgn h_i(t)$と等価である．
%
In the zero noise limit, equation (\ref{eq:E_path}) becomes 
\begin{equation}
\ll g(\vec{\sigma},\vec{v}) \gg 
= \biggl< \int {\cal D} \vec{v} \Tr{\vec{\sigma}} g(\vec{\sigma},\vec{v}) p[\sigma(0)] \prod_{s=1}^t \delta_{\sigma(s), \sgn h (\vec{\sigma},\vec{v},s-1)} \biggr>_{\vec{\xi}}, 
\label{eq:E_path2}
\end{equation}
%Note that the spin variables are deterministic 
%even if (\ref{eq:E_path2}) includes the Gaussian random fields $\vec{v}$. 
When the spin variables have periodicity as $\sigma(s+l)=\sigma(s)$, 
the Gaussian random fields are also deterministic as $v(s+l)=v(s)$ (See Appendix \ref{app:0}). 
%Hence, the spin variable evolves deterministically. 
For any function $\phi(\{\sigma(s)\})$ and any constants $c_0, \cdots ,c_{t} \in \{ -1,1 \}$, the following identity holds: 
\begin{equation}
\sum_{\sigma (0), \cdots ,\sigma (t) \in \{ -1,1 \}} \phi(\sigma(0), \cdots ,\sigma(t))
=\sum_{\sigma (0), \cdots ,\sigma (t) \in \{ -1,1 \}}  \phi(c_0 \sigma(0), \cdots , c_t \sigma(t)). 
\label{eq:const}
\end{equation}
Applying (\ref{eq:const}) to (\ref{eq:E_path2}) and substituting (\ref{eq:transformation}), we obtain 
\begin{eqnarray}
& & \ll g(\vec{\sigma},\vec{v}) \gg \nonumber \\
&=& \biggl< \int {\cal D} \vec{v} \Tr{\vec{\sigma}} g(\{c_0\sigma(0),\cdots,c_t\sigma(t)\},\vec{v}) p[c_0\sigma(0)] \prod_{s=1}^t \delta_{c_s\sigma(s), \sgn h (\{c_0\sigma(0),\cdots,c_t\sigma(t)\},\vec{v},s-1)} \biggr>_{\vec{\xi}} \nonumber \\
&=& \biggl< \int {\cal D} \vec{v} \Tr{\vec{\chi}} g(\{c_0\eta_0\chi (0),\cdots,c_t\eta_t\chi (t)\},\vec{v}) \nonumber \\
& & \qquad \times p[c_0\eta_0\chi (0)] \prod_{s=1}^t \delta_{c_s\eta_s\chi (s), \sgn h (\{c_0\eta_0\chi (0),\cdots,c_t\eta_t\chi (t)\},\vec{v},s-1)} \biggr>_{\vec{\xi}}, \nonumber \\
&=& \biggl< \int {\cal D} \vec{v} \Tr{\vec{\chi}} g(\{\chi (0),\cdots,\chi (t)\},\vec{v}) p[\chi (0)] \prod_{s=1}^t \delta_{\chi (s), \sgn h (\{\chi (0),\cdots,\chi (t)\},\vec{v},s-1)} \biggr>_{\vec{\xi}}, 
\label{eq:E_path3}
\end{eqnarray}
with 
\begin{equation}
\Tr{\vec{\chi}} \equiv \sum_{\eta_0 \chi(0), \cdots , \eta_t \chi (t) \in \{ -1,1 \} }
= \sum_{\chi(0), \cdots , \chi (t) \in \{ -1,1 \} }.
\end{equation}
In derivation of (\ref{eq:E_path3}), we put the constants $\{c_s\}$ as $c_s = \eta_s$ for all $s$. 
Generality is kept even if $c_s = \eta_s$ for all $s$. 
Namely, with respect to the transformed spin variable $\chi(s)$, 
the effective single spin described by (\ref{eq:E_path3}) is 
\begin{equation}
\chi (s) = \sgn h (\{\chi (0),\cdots,\chi (t)\},\vec{v},s-1). 
\label{eq:ess}
\end{equation}
The transformed spin variables $\chi(s)$ are deterministic 
even if (\ref{eq:ess}) includes the Gaussian random fields $\vec{v}$, 
since the Gaussian random fields are deterministic. 
In order to get rid of the self-interaction, we assume the Maxwell construction ansatz. 
Using the identity $\chi(s)=\chi(s')$ for any $s,s'$ to (\ref{eq:ess}) and applying the Maxwell construction, we get 
\begin{eqnarray}
\chi (s) 
&=& \sgn [\xi^s m(s-1)+\theta(s-1)+\sqrt{\alpha}v(s-1)+ \sum_{s'<s} \Gamma(s,s') \chi(s')] \nonumber \\
&=& \sgn [\xi^s m(s-1)+\theta(s-1)+\sqrt{\alpha}v(s-1)+ \chi (s) \sum_{s'<s} \Gamma(s,s')] \nonumber \\
&=& \sgn h(\vec{v},s-1), \label{eq:Maxwell}
\end{eqnarray}
with $h(\vec{v},s) \equiv \xi^{s+1} m(s)+\theta(s)+\sqrt{\alpha}v(s)$. 
Substituting (\ref{eq:Maxwell}) into (\ref{eq:E_path3}), we obtain 
\begin{equation}
\ll g(\vec{\sigma},\vec{v}) \gg 
= \biggl< \int {\cal D} \vec{v} \Tr{\vec{\chi}} g(\vec{\chi},\vec{v}) p[\chi (0)] \prod_{s=1}^t \delta_{\vec{\chi}, \sgn h (\vec{v},s-1)} \biggr>_{\vec{\xi}}. 
\label{eq:E_path4}
\end{equation}
Thus, we can get rid of the self-interaction in the single spin problem by using the Maxwell construction in the zero noise limit, i.e., $T=0$. 
Since (\ref{eq:E_path4}) can be factorized with respect to the tranformed spin variables $\chi(s)$ at different times, we can easily perform the spin summations. 
%(\ref{eq:pp})式は時刻に関して因数分解できるので，spin summationは簡単に計算できる．
%
After simple rescalings we arrive at 
%簡単な変数変換により(\ref{eq:spe_m})-(\ref{eq:spe_G})式は次のようになる．
%
\begin{eqnarray}
m(s) &=& \biggl< \xi^s \int {\cal D} \vec{v} \sgn h(\vec{v},s-1) \biggr>_{\vec{\xi}}, \label{eq:m2} \\
C(s,s') &=& \delta_{s,s'} +[1-\delta_{s,s'}] \biggl< \int {\cal D} \vec{v} [\sgn h(\vec{v},s-1)] [\sgn h(\vec{v},s'-1)] \biggr>_{\vec{\xi}}, \\
G(s,s') &=& \delta_{s,s'-1} \lim_{\beta \to \infty} \beta \biggl\{ 1 - \biggl< \int {\cal D} \vec{v} \tanh^2 \beta h(\vec{v},s-1) \biggl>_{\vec{\xi}} \biggr\} . \label{eq:G2}
\end{eqnarray}
%with $\{d\vec{v}d\vec{w}\}=\prod_{s<t}\frac{dv(s)dw(s)}{2\pi}$. 
%
%Note that we can also obtain (\ref{eq:m2})-(\ref{eq:G2}) from (\ref{eq:E_path4_dash}). 
We now calculate the matrix $\vec{R}$ under the condition of (\ref{eq:condition}). 
%(\ref{eq:condition})式の条件のときの$\vec{R}$を計算する．
%
Since the matrices $\vec{G}$ and $\vec{C}$ become Toeplitz matrices (especially $\vec{C}$ is symmetric) under this conditions, 
$\vec{C}$ and $\vec{G}$ can be approximately regarded as commuting matrices, i.e., $\vec{C}\vec{G}=\vec{G}\vec{C}$. 
%このとき，$\vec{G}$はテプリッツ行列，$\vec{C}$は対称テプリッツ行列となるため，$\vec{C}$と$\vec{G}$は近似的に可換$\vec{C}\vec{G}=\vec{G}\vec{C}$となる．
%
Therefore, the matrix $\vec{R}$ simplifies to 
%また，(\ref{eq:R})式の$\vec{R}=(R(s,s'))$も$R(s,s')=R(s-s')$とおくことができる．
%
\begin{eqnarray}
\vec{R}
&=& \sum_{a=0}^{l-1} \vec{G}^a \biggl( \sum_{m,n\ge 0} \vec{G}^{ml} (\vec{G}^\dagger)^{nl} \vec{C} \biggr) (\vec{G}^\dagger)^a \nonumber \\
&=& \sum_{a=0}^{l-1} \vec{G}^a \biggl( [\vec{1}-\vec{G}^l]^{-1} [\vec{1}-(\vec{G}^\dagger)^l]^{-1} \vec{C} \biggr) (\vec{G}^\dagger)^a \nonumber \\
&=& [\vec{1}-\vec{G}\vec{G}^\dagger]^{-1}[\vec{1}-(\vec{G}\vec{G}^\dagger)^l] [\vec{1}-\vec{G}^l]^{-1} [\vec{1}-(\vec{G}^\dagger)^l]^{-1} \vec{C}, 
\label{eq:R}
\end{eqnarray}
We consider the persistent parts of $C(\tau)$ and $R(\tau)$ as $C(\tau)\to q$ and $R(\tau)\to r$ for $\tau\to\infty$ and also consider the non-persistent parts $\tilde{C}(\tau)\to 0$ and $\tilde{R}(\tau)\to 0$, i.e., $C(\tau)=q+\tilde{C}(\tau)$ and $R(\tau)=r+\tilde{R}(\tau)$. 
%$\tau\to\infty$で$C(\tau)\to q$,$R(\tau)\to r$に収束するとして，$C(\tau)=q+\tilde{C}(\tau)$,$R(\tau)=r+\tilde{R}(\tau)$とおく．
%
Upon rewriting $G(\tau)=\beta\delta_{\tau,1}[1-\tilde{q}]$ and $r=q\rho$ given by (\ref{eq:R}), we obtained 
%このとき，(\ref{eq:R})式より$r=q\rho$とおける．表記を簡単にするために，$G(\tau)=\beta\delta_{\tau,1}[1-\tilde{q}]$とおくと，(\ref{eq:m2})-(\ref{eq:G2})式は次のようになる．
%
\begin{eqnarray}
m
&=& \biggl< \xi \int Dz \int \tilde{{\cal D}} \vec{v} \sgn h(\vec{v},0|z) \biggr>_{\vec{\xi}}, \\
q
&=& \biggl< \int Dz \int \tilde{{\cal D}} \vec{v} [\sgn h(\vec{v},\tau|z)][\sgn h(\vec{v},0|z)]\biggr>_{\vec{\xi}}, \\
\tilde{q}
&=& \lim_{\beta\to\infty} \biggl< \int Dz \int \tilde{{\cal D}} \vec{v} \tanh^2 \beta h(\vec{v},0|z) \biggr>_{\vec{\xi}}, \\
r
&=& q\rho,
\end{eqnarray}
from (\ref{eq:m2})-(\ref{eq:G2}) 
with $\tilde{{\cal D}} \vec{v} \equiv d\vec{v} e^{- \frac 12 \vec{v} \cdot \tilde{\vec{R}}^{-1} \vec{v}} [(2\pi )^t|\tilde{\vec{R}}|]^{-1/2}$, 
$Dz=\frac{dz}{\sqrt{2\pi}}e^{-z^2/2}$ and $h(\vec{v},\tau|z)\equiv \xi^\tau m +\theta + z\sqrt{\alpha q \rho}+\sqrt{\alpha}v(\tau)$. 
%ここで，$Dz=\frac{dz}{\sqrt{2\pi}}e^{-z^2/2}$である．
%
The matrix $\vec{G}$ is given by 
%(\ref{eq:G2})式より，$\vec{G}$は次のような行列であることがわかる．
%
\begin{equation}
\vec{G}=
\left(
\begin{array}{ccccc}
0      & 0      & 0      & \cdots & 0      \\
G(1)   & 0      & 0      & \cdots & 0      \\
0      & G(1)   & 0      & \cdots & 0      \\
\vdots &        & \ddots &        & \vdots \\
0      & \cdots & 0      & G(1)   & 0      \\
\end{array}
\right) .
\end{equation}
from (\ref{eq:G2}). 
Equation (\ref{eq:R}) can be changed to 
%(\ref{eq:R})式は次のように変形できる．
%
\begin{equation}
[\vec{1}-(\vec{G}^\dagger)^l-\vec{G}^l-(\vec{G}^\dagger\vec{G})^l] [\vec{1}-(\vec{G}\vec{G}^\dagger)^l]^{-1} [\vec{1}-\vec{G}\vec{G}^\dagger] \vec{R}=\vec{C}. 
\label{eq:R_modified}
\end{equation}
The identity $\vec{G}\vec{G}^\dagger \simeq G(1)^2\vec{1}$ holds in the case where time $s$ is sufficiently large, so the left-hand side of (\ref{eq:R_modified}) becomes 
%時刻$t$が大きいとき，$\vec{G}\vec{G}^\dagger \simeq G(1)^2\vec{1}$と近似できるので，(\ref{eq:R_modified})式の左辺は次のようになる．
%
\begin{eqnarray}
& & [\vec{1}-(\vec{G}^\dagger)^l-\vec{G}^l-(\vec{G}^\dagger\vec{G})^l] [\vec{1}-(\vec{G}\vec{G}^\dagger)^l]^{-1} [\vec{1}-\vec{G}\vec{G}^\dagger] \vec{R} \nonumber \\
&=& \left(
\begin{array}{cccc}
D(0)   & D(1)   & \cdots & D(t-1) \\
D(1)   & D(0)   & \cdots & D(t-2) \\
\vdots & \vdots & \ddots & \vdots \\
D(t-1) & D(t-2) & \cdots & D(0)   \\
\end{array}
\right) ,
\end{eqnarray}
where 
%ただし，
%
\begin{eqnarray}
D(0) &=& (1+g_1^2)g_2R(0)   -2g_1g_2R(1), \\
D(s) &=& (1+g_1^2)g_2R(s)- g_1g_2[R(s-1)+R(s+1)], \; \; (0<s<t-1)\\
D(t-1) &=& (1+g_1^2)g_2R(t-1) -2g_1g_2R(t-2), 
\end{eqnarray}
with $g_1 \equiv G(1)^l$ and $g_2 \equiv [1-G(1)^2]/[1-G(1)^{2l}]$. 
Since $\vec{D}$ and $\vec{C}$ are symmetric Toeplitz matrices, they can be diagonalized by using the discrete Fourier transformation (See Appendix \ref{app:1}). 
%行列$\vec{D},\vec{C}$は対称テプリッツ行列なので，フーリエ変換によって対角化可能である．
%
The Fourier transformations (or the Lattice Green's functions) $\hat{D}_k,\hat{C}_k$ of the matrices $\vec{D},\vec{C}$ are given by 
%$\vec{D},\vec{C}$のフーリエ変換（格子グリーン関数）$\hat{D}_k,\hat{C}_k$は次のようになる．
%
\begin{eqnarray}
\hat{D}_k &\simeq& \sum_{\tau=0}^{t-1} \biggl\{ (1+g_1^2)g_2R(\tau)-g_1g_2[R(\tau-1)-R(\tau+1)] \biggr\} e^{ik\tau}, \\
\hat{C}_k &=& \sum_{\tau=0}^{t-1} C(\tau)e^{ik\tau}. 
\end{eqnarray}
For any wave number $k$, $\hat{D}_k=\hat{C}_k$ holds when $\vec{D}=\vec{C}$. 
%$\vec{D}=\vec{C}$, のとき，任意の波数$k$について$\hat{D}_k=\hat{C}_k$となっている．
%
Taking the limit $s\to\infty$ about $\hat{D}_0=\hat{C}_0$, the following relationship is obtained: 
%$k=0$の場合について$t\to\infty$の極限をとることによって，
%
\begin{equation}
r = \frac{1-G(1)^{2l}}{(1-G(1)^2)(1-G(1)^l)^2} q. 
\label{eq:r}
\end{equation}
By working out the remaining integrals over $\vec{v}$ and setting $\theta=0$, 
we finally obtain the stationary state equations of the order parameters as follows: 
\begin{eqnarray}
m &=& {\rm erf} \biggl( \frac m{\sqrt{2\alpha \rho}} \biggr), \label{eq:sna_m} \\
U &=& \sqrt{\frac 2{\pi\alpha \rho}}e^{-\frac {m^2}{2\alpha \rho}}, \\
\rho &=& \frac{1-U^{2l}}{(1-U^2)(1-U^l)^2}, \label{eq:sna_rho}
\end{eqnarray}
with $q=1$, $\tilde{q}=1$ and $U\equiv G(1)$ where ${\rm erf}(\cdot)$ denotes Error function defined by ${\rm erf}(x) \equiv \frac 2\pi \int_0^x e^{-u^2} du$. 
%また，$q=1$,$\tilde{q}=1$であり，ここで，$U=\lim_{\beta\to\infty}G(1)$である．ここで，erf は${\rm erf}(x) \equiv \frac 2\pi \int_0^x e^{-u^2} du$ で定義される誤差関数である．
%
It turns out that these stationary state equations (\ref{eq:sna_m})-(\ref{eq:sna_rho}) given by this exact solution are equivalent 
to those of the signal-to-noise analysis (See Appendix \ref{app:2}) \cite{mimura1995}. 
%これは，出力関数を符号関数としたときの，リミットサイクルを持つ連想記憶モデルのSCSNAによる解析結果と一致する(See Appendix \ref{app:2})．
%
Figure \ref{fig:fig1} shows that the storage capacity $\alpha_c$ and the number of patterns per one limit cycle $s$. 
%記憶容量のリミットサイクル長$l$依存性を図\ref{fig:fig1}に示す．
%
Figures \ref{fig:fig2} and \ref{fig:fig3} compare the theoretical results and computer simulations for $l=3,7$ (The number of spins is $N=3000$, the number of iterations is $11$). 
The data points and error bars show the results of the computer simulation. 
With respect to the computer simulation in figure \ref{fig:fig2} and \ref{fig:fig3}, 
the stationary overlaps are defined as $m(100l)$ and $m(50l)$, respectively. 
%図\ref{fig:fig2},\ref{fig:fig3}にオーバラップの記憶率依存性について，理論値と計算機実験との比較した結果を示す．
%
It is confirmed that the theoretical results are in good agreement with the computer simulations. 
%理論は実験とよく一致していることがわかる．
%
Storage capacity monotonously increases from $\alpha_c=0.138$ ($l=1$) with the number of steps $l$ . 
In the large $l$ limit, storage capacity finally converges to $\alpha_c=0.269$, which coincides with the theoretical result of for the infinite-step sequence processing model given by D${\rm \ddot{u}}$ring et al \cite{during1998}. 
The original properties of the finite-step sequence processing model appear as long as the number of steps of a limit cycle $l$ has order $l=O(1)$. 
%また、$s=O(1)$ のオーダでのみ有限ステップの性質を持っており、
%
In the case that $l$ has the order more than $O(1)$, the properties are the same as the properties of the infinite-step sequence processing model. 
%それ以上のオーダでは無限ステップと同じ性質を持つことがわかった。
%

\section{Conclusions}
%~~~~~~~~~~~~~~~~~~~~~~~~~~~~~~~~~~~~~~~~~~~~~~~~~~~~~~~~~~~~~~~~~~~~~

We exactly analyzed an associative memory model storing an infinite number of limit cycles with finite steps by means of the path-integral method. 
%有限ステップ系列想起モデルの想起過程を経路積分によって厳密に解析した．
%
In the case where the dynamics is deterministic, 
the statistical properties are simplified like those of the infinite-step sequence processing model by using Maxwell construction. 
%温度$T=0$の決定論的ダイナミクスでの状態遷移の場合については，状態遷移の周期性を利用すると，
%各時刻での出力が持つ，過去の状態への依存性が無限ステップ系列想起モデルと同様に簡単になることがわかった．
%
We also derived the macroscopic equations for stationary states at $T=0$. 
%温度$T=0$の場合について，定常状態を記述する巨視状態方程式を想起過程から導出した．
%
\par
We obtained the dependence of storage capacity ($\alpha_c$) on the number of patterns per one limit cycle ($l$). 
%定常状態を記述する巨視状態方程式から，系列のステップ数と記憶容量の関係を求めた．
%
At $l=1$, storage capacity is $\alpha_c=0.138$, as in the Hopfield model. 
%ステップ数が$l=1$のときは，記憶容量はHopfieldモデルと同じ$\alpha_c=0.138$となる．
%
A storage capacity monotonously increases with the number of limit cycles, and converges to $\alpha_c=0.269$ at $l\simeq 10$. 
%ステップ数の増加とともに単調に記憶容量は増加し，
%ステップ数$l=10$程度で，すぐに無限ステップ系列想起モデルの記憶容量$\alpha_c=0.269$に収束する．
%
The original properties of the finite-step sequence processing model appear as long as the number of steps of the limit cycle has order $l=O(1)$. 
%ステップ数$l=O(1)$でのみ有限ステップに特有の性質を持つことがわかった．
%
\par
%According to Bolle et al, the transient dynamics given by signal-to-noise analysis is exact up to only time step 3 and inexact to step 4 or beyond. 
%Bolleらの報告によると，
%S/N解析による想起過程の解析結果は，初期の数ステップの過渡特性についても厳密解となっている．
%
In the case where the dynamics is deterministic, we also derived the stationary state equations by using Maxwell construction. 
The stationary state equations are equivalent to those of the signal-to-noise analysis. 
%また，この定常状態を記述する巨視状態方程式は，以前我々が求めたS/N解析による解析結果と一致した．
%
%---DELETED---The path-integral method, which does not require a Gaussian approximation for crosstalk noise, 
%---DELETED---and the signal-to-noise analysis, which needs such an approximation, gave the same results in the stationary state. 
%このように，S/N解析ではクロストークノイズがガウス分布に従うと仮定するにもかかわらず，
%定常状態においてはガウス近似が不必要な経路積分による解析結果と一致した．
%
This means that the signal-to-noise analysis applied to the stationary state is exact 
in spite of including errors in the middle of the transient dynamics in the zero noise limit. 
%つまり，S/N解析による定常状態の解析結果が厳密解であることがわかった．
%
%---DELETED---\par
%---DELETED---In the finite-step sequence processing model, 
%---DELETED---although the signal-to-noise analysis applied to the transient dynamics includes errors in the middle of the transitional state, 
%---DELETED---the signal-to-noise analysis can give exact solutions in both the stationary state and the first few time steps of the transitional dynamics in the case where $T=0$. 
%これにより，統計神経力学などのS/N解析による自己想起モデルや有限ステップ系列想起モデルの動特性の解析は，
%想起過程の途中の段階では誤差を含むものの，初期の数ステップと定常状態に関しては経路積分と同様に厳密であることがわかる．
%

\begin{center}
{\small \bf Acknowledgement}
\end{center}
%~~~~~~~~~~~~~~~~~~~~~~~~~~~~~~~~~~~~~~~~~~~~~~~~~~~~~~~~~~~~~~~~~~~~~

This work was partially supported by a Grant-in-Aid 
for Scientific Research on Priority Areas No. 14084212, 
for Scientific Research (C) No. 14580438, 
and for Encouragement of Young Scientists (B) No. 15700141 
from the Ministry of Education, Culture, Sports, Science and Technology of Japan.

\appendix

\section{Derivation of order parameter equations\label{app:-1}}
%~~~~~~~~~~~~~~~~~~~~~~~~~~~~~~~~~~~~~~~~~~~~~~~~~~~~~~~~~~~~~~~~~~~~~

D${\rm \ddot{u}}$ring et al discussed the sequential associative memory model by using the path-integral method \cite{during1998}. 
%D${\rm \ddot{u}}$ringらは，系列想起モデルを経路積分法によって議論している．
%
Here , we discuss for the model with finite temperature $T$, according to their paper. 
%ここでは，彼らに従い有限温度モデルに対する巨視状態方程式を示す．次の式の導出の詳細は文献\cite{during1998}を参照されたい．経路積分法によって解析する\cite{during1998,kawamura2002}．
%
Most of the technical detail to derive order parameter equations is almost identical to the paper of D${\rm \ddot{u}}$ring et al \cite{during1998}. %---ADDED---
%巨視状態方程式の導出方法はD"uringの論文とほとんど同じです．
%
%The detailed derivation is available in their paper \cite{during1998}. 
%想起過程を記述する巨視的状態方程式を導出して，その定常状態を議論する．
%

%---DELETED---\subsection{Derivation of saddle-point equations}
%~~~~~~~~~~~~~~~~~~~~~~~~~~~~~~~~~~~~~~~~~~~~~~~~~~~~~~~~~~~~~~~~~~~~~

%The dynamics (\ref{eq:dynamics}) is a Markov chain, so the path probabilities $p[\vec{\sigma}(0),\cdots ,\vec{\sigma}(t)]$ are simply given by products of the individual transition probabilities $W[\vec{\sigma}(s+1)|\vec{\sigma}(s)]$, i.e., $p[\vec{\sigma}(0),\cdots ,\vec{\sigma}(t)] = p[\vec{\sigma}(0)]\prod_{s=0}^{t-1}W[\vec{\sigma}(s+1)|\vec{\sigma}(s)]$.  %<----------(\ref{eq:dynamics})式のダイナミクスはマルコフ連鎖であるため，path probability p は個々の状態遷移確率 w の積で与えられる．
The generating functional $\bar{Z}[\vec{\psi}]$ contains both condensed and non-condensed patterns. 
We isolate the non-condensed ones by introducing the local field $\vec{h}$ and the variables $\vec{x},\vec{y}$: %<----------生成関数にデルタ関数の積分形，$\delta (x) = \int_{-\infty}^\infty \frac{d\hat{x}}{2\pi} e^{i\hat{x}x}$を用いて表現して生成関数に取り込む．%また，ある時刻には１つのパターンのみが想起されているとして，非想起パターンに関する補助変数$x_{\nu\mu}(s),y_{\nu\mu}(s)$をデルタ関数を使って導入する．
\begin{eqnarray}%(introduction of h,x,y)
1 &=& \int \frac{d\vec{h} d\hat{\vec{h}}}{(2\pi)^{Nt}} \prod_i e^{i \hat{h}_i(s) \bigl[ h_i(s)-\sum_j J_{ij} \sigma_j(s) - \theta_i(s) \bigr] } , \label{eq:def_h} \\
1 &=& \int \frac{d\vec{x} d\hat{\vec{x}}}{(2\pi)^{(p-1)t}} e^{i \sum_{s<t} \sum_{\nu =1}^{p/l} \sum_{\mu =1 (\mu \ne s \; {\rm at} \; \nu=1)}^{l} \hat{x}_{\nu\mu}(s) \bigl[ x_{\nu\mu}(s) - \frac 1{\sqrt{N}} \sum_i \xi_i^{\nu,\mu +1} \hat{h}_i(s) \bigr]} , \\
1 &=& \int \frac{d\vec{y} d\hat{\vec{y}}}{(2\pi)^{(p-1)t}} e^{i \sum_{s<t} \sum_{\nu =1}^{p/l} \sum_{\mu =1 (\mu \ne s \; {\rm at} \; \nu=1)}^{l} \hat{y}_{\nu\mu}(s) \bigl[ y_{\nu\mu}(s) - \frac 1{\sqrt{N}} \sum_i \xi_i^{\nu,\mu}    \sigma_i(s)  \bigr]} . 
\end{eqnarray}
We isolate the various relevant macroscopic observables by inserting integrals over appropriate $\delta$-functions: %<----------さらに，巨視変数を以下のように導入する．
\begin{eqnarray}%(introduction of m,k,q,Q,K)
1 &=& \int \frac{d\vec{m} d\hat{\vec{m}}}{(2\pi /N)^t}     e^{i \sum_{s<t}    \hat{m}(s)    \bigl[ m(s)    - \frac 1N \sum_i \xi_i^{1,s}   \sigma_i(s)   \bigr]} , \\
1 &=& \int \frac{d\vec{k} d\hat{\vec{k}}}{(2\pi /N)^t}     e^{i \sum_{s<t}    \hat{k}(s)    \bigl[ k(s)    - \frac 1N \sum_i \xi_i^{1,s+1} \hat{h}_i(s)  \bigr]} , \\
1 &=& \int \frac{d\vec{q} d\hat{\vec{q}}}{(2\pi /N)^{t^2}} e^{i \sum_{s,s'<t} \hat{q}(s,s') \bigl[ q(s,s') - \frac 1N \sum_i \sigma_i(s)   \sigma_i(s')  \bigr]} , \\
1 &=& \int \frac{d\vec{Q} d\hat{\vec{Q}}}{(2\pi /N)^{t^2}} e^{i \sum_{s,s'<t} \hat{Q}(s,s') \bigl[ Q(s,s') - \frac 1N \sum_i \hat{h}_i(s)  \hat{h}_i(s') \bigr]} , \\
1 &=& \int \frac{d\vec{K} d\hat{\vec{K}}}{(2\pi /N)^{t^2}} e^{i \sum_{s,s'<t} \hat{K}(s,s') \bigl[ K(s,s') - \frac 1N \sum_i \sigma_i(s)   \hat{h}_i(s') \bigr]} . \label{eq:def_K}
\end{eqnarray}
The generating functional which for $N\to\infty$ will be dominated by saddle points. 
We obtain 
\begin{equation}
\bar{Z}[\vec{\psi}] = \int d\vec{m}d\hat{\vec{m}} d\vec{k}d\hat{\vec{k}} d\vec{q}d\hat{\vec{q}} d\vec{Q}d\hat{\vec{Q}} d\vec{K}d\hat{\vec{K}} e^{N(\Psi + \Phi + \Omega)+{\cal O}(N^{1/2})}, 
\label{eq:Zbar}
\end{equation}
by substituting (\ref{eq:def_h})-(\ref{eq:def_K}) into (\ref{eq:def_Z}), where 
\begin{eqnarray}
\Psi 
&=& i \sum_{s<t} \biggl[ \hat{m}(s)m(s) + \hat{k}(s)k(s) - m(s)k(s) \biggr] \nonumber \\
& & + i \sum_{s,s'<t} \biggl[ \hat{q}(s,s')q(s,s') + \hat{Q}(s,s')Q(s,s') + \hat{K}(s,s')K(s,s') \biggr] , \label{eq:Psi} \\
\Phi
&=& \frac 1N \sum_i \ln \Tr{\vec{\sigma}} p_i(\sigma(0)) \int \{ d\vec{h} d\hat{\vec{h}} \} \nonumber \\
& & \times e^{\sum_{s<t} \bigl[ \beta \sigma(s+1) - \ln 2 \cosh \beta h(s) \bigr] } \nonumber \\
& & \times e^{- i \sum_{s,s'<t} \bigl[ \hat{q}(s,s')\sigma(s)\sigma(s') + \hat{Q}(s,s')\hat{h}(s)\hat{h}(s') + \hat{K}(s,s')\sigma(s)\hat{h}(s') \bigr] } \nonumber \\
& & \times e^{i \sum_{s<t} \hat{h}(s) \bigl[ h(s) -\theta_i(s) -\hat{k}(s)\xi_i^{1,s+1} \bigr] - i \sum_{s<t} \sigma(s) \bigl[ \hat{m}(s)\xi_i^{1,s} +\psi_i(s) \bigr] }, \label{eq:Phi} \\
\Omega 
&=& \frac 1N \ln \int \frac{d\vec{u}d\vec{v}}{(2\pi)^{(p-t)t}} e^{i \sum_{\mu>t} \sum_{s<t} u_{\nu,\mu+1}(s) v_{\nu,\mu}(s) } \nonumber \\
& & \times e^{-\frac 12 \sum_{\nu >1,\mu \ge 1} \sum_{s,s'<t} \bigl[ u_{\nu,\mu}(s)Q(s,s')v_{\nu,\mu}(s') + u_{\nu,\mu}(s)K(s,s')v_{\nu,\mu}(s') \bigr] } \nonumber \\
& & \times e^{-\frac 12 \sum_{\nu >1,\mu \ge 1} \sum_{s,s'<t} \bigl[ v_{\nu,\mu}(s)K(s,s')u_{\nu,\mu}(s') + v_{\nu,\mu}(s)q(s,s')v_{\nu,\mu}(s') \bigr] }, \label{eq:Omega}
\end{eqnarray}
with the shorthand $\{ d\vec{h} d\hat{\vec{h}} \} = \prod_i \frac{dh_i(s)d\hat{h}_i(s)}{2\pi}$.

%---削除---\subsection{The saddle-point equations}
%~~~~~~~~~~~~~~~~~~~~~~~~~~~~~~~~~~~~~~~~~~~~~~~~~~~~~~~~~~~~~~~~~~~~~

In the limit $N\to\infty$, the integral (\ref{eq:Zbar}) will be dominated by saddle point of the extensive exponent $\Psi+\Phi+\Omega$. %<----------(\ref{eq:Zbar})式の積分値は，$N\to\infty$の極限で指数部分$\Psi+\Phi+\Omega$の極値によって支配される．そこで，$N\to\infty$として鞍点法でこの積分値を評価する．
The saddle-point equations which are derived by differentiation with respect to integration variables $\{ \vec{m},\hat{\vec{m}},\vec{k},\hat{\vec{k}},\vec{q},\hat{\vec{q}},\vec{Q},\hat{\vec{Q}},\vec{K},\hat{\vec{K}} \}$ are as follows: %<----------まず，鞍点方程式を導出する．積分変数$\{ \vec{m},\hat{\vec{m}},\vec{k},\hat{\vec{k}},\vec{q},\hat{\vec{q}},\vec{Q},\hat{\vec{Q}},\vec{K},\hat{\vec{K}} \}$について指数部分を微分して鞍点を求めると，$s,s'<t$に対して次のようになる．
\begin{eqnarray}
\hat{m}(s)&=&k(s)=0, \\
Q(s,s')&=&\hat{q}(s,s')=0, \\
m(s)&=&\hat{k}(s)=\lim_{N\to\infty}\frac 1N\sum_i<\sigma(s) \xi_i^{1,s}>_i, \label{eq:sp_m} \\
q(s,s')&=&C(s,s')=\lim_{N\to\infty}\frac 1N\sum_i <\sigma(s)\sigma(s')>_i, \label{eq:sp_C} \\
K(s,s')&=&iG(s,s')= i \lim_{N\to\infty} \frac 1N \sum_i \frac{\partial <\sigma(s)>_i}{\partial \theta_i(s')}, \label{eq:sp_G} \\
\hat{Q}(s,s')&=&\biggl. i \lim_{\vec{Q}\to\vec{0}} \frac{\partial \Omega}{\partial K(s,s')} \biggr|_{\rm saddle}, \label{eq:sp_Qhat} \\
\hat{K}(s,s')&=&\biggl. \frac{\partial \Omega}{\partial K(s,s')} \biggr|_{\rm saddle}, \label{eq:sp_Khat} \\
\end{eqnarray}
where $f|_{\rm saddle}$ denotes an evaluation of a function $f$ at the dominating saddle-point, $<\cdot>_i$ denotes 
\begin{eqnarray}
<f(\vec{\sigma},\vec{h},\hat{\vec{h}})>_i
&=& \biggl< \frac{\Tr{\vec{\sigma}} \int \{ d\vec{h} d\hat{\vec{h}} \} W_i(\vec{\sigma},\vec{h},\hat{\vec{h}})f(\vec{\sigma},\vec{h},\hat{\vec{h}})}
        {\Tr{\vec{\sigma}} \int \{ d\vec{h} d\hat{\vec{h}} \} W_i(\vec{\sigma},\vec{h},\hat{\vec{h}})} \biggr>_{\vec{\xi}}, \\
W_i(\vec{\sigma},\vec{h},\hat{\vec{h}})
&=& p_i(\sigma(0)) \biggl[ e^{\sum_{s<t} \bigl( \beta\sigma(s+1)h(s)-\ln 2\cosh \beta h(s) \bigr)  } \nonumber \\
& & \times e^{ i\sum_{s<t} \bigl( \hat{h}(s) \{ h(s)-\theta_i(s)-\hat{k}(s)\xi_i^{1,s+1} \} - \theta(s)\hat{m}(s)\xi_i^{1,s} \bigr)  } \nonumber \\
& & \times e^{-i\sum_{s,s'<t} \bigl( \hat{q}(s,s')\sigma(s)\sigma(s')+\hat{Q}(s,s')\hat{h}(s)\hat{h}(s')+\hat{K}(s,s')\sigma(s)\hat{h}(s') \bigr) } \biggr], 
\end{eqnarray}
and $<\cdot>_{\vec{\xi}}$ denotes the average over the condensed patterns. 
We now calculate the right-hand sides of (\ref{eq:sp_Qhat}) and (\ref{eq:sp_Khat}). 
The eigenvalues $s_\mu$ of a matrix $\vec{S}$ are given by $s_\mu =e^{2\pi i\mu /l}$ (the multiplicity is $p/l$) from $|\lambda \vec{1} - \vec{S}| = |\lambda \vec{1} - \vec{S}'|^{p/l}$. 
Since the matrix $\vec{S}'$ is a unitary matrix, i.e., $\vec{S}'^\dagger\vec{S}'=\vec{1}$, the matrix $\vec{S}$ is also unitary, $\vec{S}^\dagger\vec{S}= {\rm diag}(\vec{S}'^\dagger,\cdots,\vec{S}'^\dagger){\rm diag}(\vec{S}',\cdots,\vec{S}') = {\rm diag}(\vec{S}'^\dagger\vec{S}',\cdots,\vec{S}'^\dagger\vec{S}') = \vec{1}$. 
A $(\mu,\mu)$-element of $(\vec{S^\dagger})^m\vec{S}^n$ becomes $[(\vec{S^\dagger})^m\vec{S}^n]_{\mu\mu} =\delta_{mn}$ by using the unitarity of $\vec{S}$, where $\delta_{mn}$ denotes Kronecker's delta function. 
The identity $(\vec{S})^l=(\vec{S}^\dagger)^l=\vec{S}$ is established because $(\vec{S}')^l=(\vec{S}'^\dagger)^l=\vec{S}'$. 
Therefore, the identity $[(\vec{S^\dagger})^m\vec{S}^n]_{\mu\mu}=\delta_{mn}$ holds for the following value $(m,n)$: 
\begin{equation}
\left\{
\begin{array} {l}
m=m'l+a, \\
n=n'l+a,
\end{array} \right. 
\end{equation}
with $m',n' \in \{0,1,\cdots \}$ and $a \in \{ 0,\cdots,l-1 \}$. 
The multiplicity of the eigenvalues of the matrix $\vec{S}'$ is 1, so ${\rm rank}(\vec{S}'-s_\mu\vec{1})=l-1$. 
Hence, ${\rm rank}(\vec{S}-s_\mu \vec{1})= (p/l) {\rm rank}(\vec{S}'-s_\mu \vec{1}) = p-p/l$. 
%---DELETE---This means that the dimension of the eigenspace with the eigenvalue $s_\mu$ is equal to the multiplicity of eigenvalue of $s_\mu$ for any $\mu$. 
%というように，任意の$\mu$について固有値$s_\mu$に対する$\vec{S}$の固有空間の次元が固有値$s_\mu$の重複度に一致するので，適当な正則行列$\vec{P}$によって$\vec{S}$は次のように対角化可能である．
%
This means that the matrix $\vec{S}$ can be diagonalized by an appropriate non-singular matrix as 
%となるので，適当な正則行列$\vec{P}$によって$\vec{S}$は次のように対角化可能である．%(任意の$\mu$について固有値$s_\mu$に対する$\vec{S}$の固有空間の次元が固有値$s_\mu$の重複度に一致する)%<----------となるので，適当な正則行列$\vec{P}$によって$\vec{S}$は次のように対角化可能である．
%
\[
{\rm diag} (\overbrace{s_0,\cdots,s_0}^{p/l},\cdots,\overbrace{s_{l-1},\cdots,s_{l-1}}^{p/l}). 
\]
By working out the saddle-point equation (\ref{eq:sp_Qhat}), $\hat{\vec{Q}}$ becomes as follows: 
\begin{equation}
\hat{Q}(s,s')=-\frac 12 \alpha i\sum_{m,n\ge 0}\lim_{p\to\infty}\sum_{\mu\le p}\{ (\vec{S}\otimes\vec{G})^m[\vec{1}\otimes\vec{C}](\vec{S}^\dagger\otimes\vec{G}^\dagger)^n \}_{\mu\mu}(s',s). 
\end{equation}
Hence $\hat{\vec{Q}}$ is given by 
\begin{equation}
\hat{\vec{Q}} = -\frac 12 \alpha i \sum_{a=0}^{l-1} \sum_{m',n'\ge 0}\vec{G}^{m'l+a} \vec{C}(\vec{G}^\dagger)^{n'l+a}. 
\end{equation}
We define a matrix $\vec{\Gamma}=\vec{S}\otimes\vec{R}$ as having matrix elements $\Gamma_{\mu\mu '}(s,s')=S_{\mu\mu '}R(s,s')$ for $\mu,\mu' \in \{1,\cdots,p \}$ and $s,s' \in \{ 0,\cdots,t-1 \}$ where $\vec{y}=\vec{\Gamma\vec{x}}$ will operate as $y_\mu(s)=\sum_{\mu '>t}\sum_{s'<t}S_{\mu\mu '}R(s,s')x_{\mu '}(s')$ for each $(\mu,s)$. 
Equation (\ref{eq:sp_Khat}) reduces to 
\begin{eqnarray}
\hat{K}(s,s')
&=& -\frac 12 \alpha \frac{\partial}{\partial G(s,s')} \lim_{p\to\infty} \frac 1p \{ \ln\det [\vec{1}\otimes\vec{1}-\vec{S}^\dagger\otimes\vec{G}^\dagger ] + \ln\det [\vec{1}\otimes\vec{1}-\vec{S}\otimes\vec{G}] \} \nonumber \\
&=& -\alpha \frac{\partial}{\partial G(s,s')} \lim_{p\to\infty} \frac 1p \ln \prod_{\mu=0}^{l-1} \biggl( \det [\vec{1}-e^{ 2\pi i \mu /l}\vec{G}^\dagger ] \biggr)^{p/l} \\
\hat{\vec{K}}
&=&\frac{\alpha}l \sum_{\mu=0}^{l-1} e^{2\pi i\mu /l} [\vec{1}-e^{2\pi i\mu /l}\vec{G}^\dagger]^{-1}. 
\end{eqnarray}
Replacing $\xi_i^{1,s}\to\xi^{s}$, we obtain the order parameters of (\ref{eq:spe_m})-(\ref{eq:spe_G}).

\section{The periodicity of the Gaussian random field \label{app:0}}
%~~~~~~~~~~~~~~~~~~~~~~~~~~~~~~~~~~~~~~~~~~~~~~~~~~~~~~~~~~~~~~~~~~~~~

Using time-translation invariant ansatz, i.e., $C(s,s')=C(s-s')$, $C(s)$ also have periodicity as 
\begin{equation}
C(s+l)=C(s), 
\end{equation}
for any $s$, when spin variables $\sigma(s)$ have periodicity, i.e., $\sigma(s+l)=\sigma(s)$. 
It is confirmed that the vector $\vec{v}$ obeys the Gassian distribution 
with mean $\vec{0}$ and covariance matrix $\vec{R}$ from equation (\ref{eq:E_path}). 
The covariance matrix $\vec{R}$ of the Gaussian random fields $\vec{v}$ is given by (\ref{eq:R}). 
Since the matrix $\vec{R}$ is also symmetric Toeplitz matrix, we can put $R(s,s')=R(s-s')$ where $R(s,s')$ are the elements of the matrix $\vec{R}$. 
When $C(s+l)=C(s)$, we approximately get $R(s+l)=R(s)$. 
The correlation coefficient between cross-talk noise $v(s)$ and $v(s+l)$ becomes 
\[
{\rm Corr}(v(s+l),v(s))
=\frac{{\rm Cov}(v(s+l),v(s))}{\sqrt{V(v(s+l))V(v(s))}}
=\frac{R(s+l,s)}{\sqrt{R(s+l,s+l)R(s,s)}}
=1, 
\]
for any $s,s'$. 
The $v(s)$ distribution and the $v(s+l)$ distribution have the same mean and variance, i.e., $E(v(s))=E(v(s+l))=0$ and $V(v(s))=R(s,s)=R(0)=R(s+l,s+l)=V(v(s+l))$. 
Therefore, the identity 
\begin{equation}
v(s+l)=v(s), 
\end{equation}
holds for any $s$. 
Hence, the Gaussian random fields $v(s)$ also have periodicity as $v(s+l)=v(s)$ when spin variables have periodicity as $\sigma(s+l)=\sigma(s)$. 
Therefore, $v(s)$ can be considered as quenched noise. 
After the states $v(0), \cdots , v(l-1)$ occur at random, $v(s)$ continues taking the same values periodically. 
Therefore, we can assume that $v(s)$ is deterministic for a fixed site. 
The Gaussian random fields $v(s)$ are only distributed with respect to site index $i$.

\section{The Fourier transformation of the symmetric Toeplitz matrix \label{app:1}}
%~~~~~~~~~~~~~~~~~~~~~~~~~~~~~~~~~~~~~~~~~~~~~~~~~~~~~~~~~~~~~~~~~~~~~

Symmetric Toeplitz matrices can be diagonalized by using the discrete Fourier transformation. 
%対称テプリッツ行列はフーリエ変換によって対角化できる．
%
The Fourier transformation was used to obtain the identity of (\ref{eq:r}). 
%ここでは，(\ref{eq:r})式を求めるためにフーリエ変換を用いている．
%
The Fourier transformation is defined as 
%フーリエ変換を次式で定義する．
%
\begin{equation}
\hat{\xi}_k = \frac 1{\sqrt{t}} \sum_j \xi_j e^{-ikj}, 
\end{equation}
and the inverse Fourier transformation defined as 
%次に，逆フーリエ変換を次式で定義する．
%
\begin{equation}
\xi_j = \frac 1{\sqrt{t}} \sum_k \hat{\xi}_k e^{ikj},
\end{equation}
where $k$ denotes wave number and its degree of freedom is $t$. 
%ここで，$k$は波数を示し，その自由度は$t$である．
%
Each component of $k$ takes the value $0,\frac{2}{t}\pi,\frac{4}{t}\pi, \cdots,\frac{2(t-1)}{t}\pi$.
%$k$の各要素は$0,\frac{2}{t}\pi,\frac{4}{t}\pi, \cdots,\frac{2(t-1)}{t}\pi$の値をとる．
%
The following symmetric Toeplitz matrix can be diagonalized by using the Fourier representation: 
%次の対称テプリッツ行列$\vec{D}$，
%
\begin{equation}
\vec{D}=
\left(
\begin{array}{cccc}
D_0     & D_1     & \cdots & D_{t-1} \\
D_1     & D_0     & \cdots & D_{t-2} \\
\vdots  & \vdots  & \ddots & \vdots  \\
D_{t-1} & D_{t-2} & \cdots & D_0     \\
\end{array}
\right) . 
\end{equation}
The Fourier representation of the quadratic form $\vec{\xi}^T \vec{D} \vec{\xi}$ becomes 
%の2次形式$\vec{\xi}^T \vec{D} \vec{\xi}$のフーリエ表記を求める．
%
\begin{eqnarray}
\vec{\xi}^T \vec{D} \vec{\xi}
&=& \sum_{i=1}^t \sum_{j=1}^t \xi_i D_{|i-j|} \xi_j \nonumber \\
&=& \sum_{\tau=0}^{t-1} \sum_{j=1}^t \xi_j D_\tau \xi_{j-\tau} \nonumber \\
&=& \sum_{\tau} \sum_j \biggl( \frac 1{\sqrt{t}} \sum_{k_1} \hat{\xi}_{k_1} e^{ik_1j} \biggr)  D_\tau \biggl( \frac 1{\sqrt{t}} \sum_{k_2} \hat{\xi}_{k_2} e^{ik_2(j-\tau)} \biggr) \nonumber \\
&=& \sum_k \hat{\xi}_k \biggl( \sum_{\tau} D_\tau e^{ik\tau} \biggr) \hat{\xi}_{-k}, 
\end{eqnarray}
where the index $j$ of the variables $\xi_j$ is understood to be taken modulo $t$. 
%ここで，$\xi_j$の添え字は$ \mod t$をとるとする．
%
Therefore, the Fourier transformation of the symmetric Toeplitz matrix $\vec{D}=(D_\tau)$ is given by 
%よって，対称テプリッツ行列$\vec{D}=(D_\tau)$のフーリエ変換$\hat{\vec{D}}=(\hat{D}_k)$は，
%
\begin{equation}
\hat{D}_k=\sum_{\tau} D_\tau e^{ik\tau}. 
\end{equation}
%となる．
This transformation is also called the lattice Green's function. 
%また，この関数は格子グリーン関数とも呼ばれる．
%

\section{The signal-to-noise analysis of the finite-step sequence processing model \label{app:2}}
%~~~~~~~~~~~~~~~~~~~~~~~~~~~~~~~~~~~~~~~~~~~~~~~~~~~~~~~~~~~~~~~~~~~~~

We discuss the stability of limit cycles by means of the signal-to-noise analysis. 
Let us consider the following deterministic synchronous dynamics: 
\begin{equation}
x_i^{t+1} = F ( \sum_{j=1}^N J_{ij} x_j^t ) , \label{eq:eq-st}
\end{equation}
where the $x_i^t$ represents the state of the $i$-th neuron at time $t$ and $F(\cdot)$ denotes an output function. 
The retrieval state converges to some limit cycle with $l$ steps. 
We introduced the Poincar\'{e} map to get the states every $l$ steps in the steady state. 
The periodic state can be transformed into a stable state by using this map. 
Hence, we can discuss the properties of a stability of limit cycles. 

Let us consider the case of convergence to periodic states of the limit cycle retrieval. 
We assume $x_i^t = x_i^{t-l}$ in (\ref{eq:eq-st}). 
The overlap between the $\mu$-th memory pattern $\vec{\xi}^{\nu \mu}$ of the $\nu$-th limit cycle and the network state $\vec{x}$ is defined as 
\begin{equation}
    m_{\nu \mu}^t = \frac 1N \sum_{i=1}^N \xi_i^{\nu \mu} x_i^t. 
    \label{eq:overlap-t}
\end{equation}
The dynamics (\ref{eq:eq-st}) can be rewritten as 
\begin{eqnarray}
    x_i^{t+1} &=& F(h_i^t) \label{eq:eq-st-t} \\
    h_i^t
    &=& \sum_{j=1}^N J_{ij} x_j^t
     =  \sum_{\nu=1}^{p/l} \sum_\mu \xi_i^{\nu \mu +1} m_{\nu \mu}^t, 
    \label{eq:hi-t1}
\end{eqnarray}
where $h_i^t$ is a local field. 
Now let us consider the case to retrieve the 1st limit cycle $\mbox{\boldmath $\xi$}^{1 \mu}$, $\mu \in \{ 1, \cdots ,l \}$. 
We assume the memorized pattern of another limit cycles $\mbox{\boldmath $\xi$}^{\nu \mu}, \nu \ne 1$ does not have a finite overlap. 
We consider retrieval solutions in which $m_{1 \mu}^t \sim O(1)$, and $m_{\nu \mu}^t \sim O( 1 / \sqrt{N} ), \nu \geq 2 $. 
We assume that the components $x_i$ of the equilibrium state $\vec{x}$ are independent on the unit number $i$ in the limit $N \to \infty$. 
It is necessary to assume that the self-averaging property to holds so that the site average can be replaced by an average over the random patterns and random variable $\vec{x}$. 
In this situation the overlap $m_{\nu\mu}^t$ need not to be a random variable.
\begin{eqnarray}
    m_{\nu \mu}^t &=& \bar{m}_{\nu \mu}^t + U_t m_{\nu \mu}^{t-1} ,
    \label{eq:overlap-t2} \\
    \bar{m}_{\nu \mu}^t &=& \frac 1N \sum_i \xi_i^{\nu ,\mu}
                            x_i^{t (\nu \mu )}, \label{eq:m-bar} \\
    U_t &=& \frac 1N \sum_i x_i^{'t (\nu \mu )},
\end{eqnarray}
where
\begin{eqnarray}
   x_i^{t(\nu \mu )}  &=& F ( \sum_{(\nu ' \mu ') \ne (\nu ,\mu -1)}
                              \xi^{\nu ' \mu '+1} m_{\nu ' ,\mu '}^{t-1} )
\\
   x_i^{'t(\nu \mu )} &=& F'( \sum_{(\nu ' \mu ') \ne (\nu ,\mu -1)}
                              \xi^{\nu ' \mu '+1} m_{\nu ' \mu '}^{t-1} ) .
\end{eqnarray}
By applying (\ref{eq:overlap-t2}) repeatedly, we obtain 
\begin{eqnarray}
    m_{\nu \mu}^t
    &=& \bar{m}_{\nu \mu}^t
      + U_t         \bar{m}_{\nu \mu -1}^{t-1}
      + U_t U_{t-1} \bar{m}_{\nu \mu -2}^{t-2} + \cdots \nonumber \\
    & & \; \; \; \;
      + U_t U_{t-1} \cdots U_{t-l+2} \bar{m}_{\nu \mu -l+1}^{t-l+1}
      + U_t U_{t-1} \cdots U_{t-l+1} m_{\nu \mu -l}^{t-l}. 
\end{eqnarray}
Since the retrieval state is assumed to be steady, i.e., $x_i^{t-l}=x_i^t $, the identity $m_{\nu \mu -l}^{t-l} = m_{\nu \mu}^t$ holds. 
Therefore the overlap becomes 
\begin{equation}
    m_{\nu \mu} ^t
    = (1-\prod_{k=0}^{l-1}U_{t-k})^{-1} \left[
       \bar{m}_{\nu \mu}^t
       + \sum_{k=1}^{l-1}\prod_{k'=0}^{k-1} U_{t-k'} \bar{m}_{\nu
\mu -k}^{t-k}
       \right] .
    \label{eq:overlap-t3}
\end{equation}
Substituting (\ref{eq:m-bar}) into (\ref{eq:overlap-t3}), we obtain 
\begin{eqnarray}
    m_{\nu \mu} ^t
    &=& \frac 1N (1-\prod_{k=0}^{l-1}U_{t-k})^{-1} \left[
        \sum_{j=1}^N \xi_j^{\nu \mu} x_j^{t (\nu \mu )} \right.
        \nonumber \\
    & & \; \; \; \; \;
      + \left.
        \sum_{j=1}^N \xi_j^{\nu ,\mu} x_j^{t (\nu \mu )}
        \sum_{k=1}^{l-1} ( \prod_{k'=0}^{k-1} U_{t-k'} )
        \sum_{j=1}^N \xi_j^{\nu ,\mu -k} x_j^{t-k (\nu \mu -k)}
        \right] ,
    \label{eq:overlap-t4}
\end{eqnarray}
Replacing $\xi_i^{1 \mu} \to \xi_i^\mu$ and $m_{1 \mu}^t \to m_\mu^t $, and substituting (\ref{eq:overlap-t4}) into (\ref{eq:hi-t1}), the local field $h_i^t$ can be rewritten as 
\begin{eqnarray}
    h_i^t
    &=& \sum_\mu \xi_i^{\mu +1} m_\mu^t \nonumber \\
    & & \; \; \; \; \;
      + \frac 1N (1-\prod_{k=0}^{l-1}U_{t-k})^{-1} \left[
        \sum_{\nu \geq 2}^{\alpha N /l} \sum_{\mu =1}^l
        \xi_i^{\nu \mu +1} \xi_i^{\nu \mu } x_i^{t (\nu \mu )} \right.
        \nonumber \\
    & & \; \; \; \; \; \; \; \; \; \;
      + \left.
        \sum_{k=0}^{l-1} ( \prod_{k'=0}^{k-1} U_{t-k'} )
        \sum_{\nu \geq 2}^{\alpha N /l} \sum_{\mu =1}^l
        \xi_i^{\nu \mu +1} \xi_i^{\nu \mu -k} x_i^{t-k (\nu \mu -k)}
        \right]
        \nonumber \\
    & & \; \; \; \; \;
      + \frac 1N (1-\prod_{k=0}^{l-1}U_{t-k})^{-1} \left[
        \sum_{\nu \geq 2}^{\alpha N /l} \sum_{\mu =1}^l \sum_{j \ne i}^N
        \xi_i^{\nu \mu +1} \xi_j^{\nu \mu } x_j^{t (\nu \mu)} \right.
        \nonumber \\
    & & \; \; \; \; \; \; \; \; \; \;
      + \left.
        \sum_{k=0}^{l-1} ( \prod_{k'=0}^{k-1} U_{t-k'} )
        \sum_{\nu \geq 2}^{\alpha N /l} \sum_{\mu =1}^l \sum_{j \ne i}^N
        \xi_i^{\nu \mu +1} \xi_j^{\nu \mu -k} x_j^{t-k (\nu \mu -k)}
        \right] .
    \label{eq:hi-t2}
\end{eqnarray}
The first term in (\ref{eq:hi-t2}) is regarded as the signal. 
The second and the third terms are regarded as the mean and the variance of the crosstalk noise, respectively. 
In the second term, if the suffix is $\mu + 1 \equiv \mu -k \, (\mod l)$, $\xi_i^{\nu \mu +1}$ and $\xi_j^{\nu \mu -k}$ are represented the same patterns.
Since these terms consist of uncorrelated random variables with the order $O(1/\sqrt{N})$, the other terms in the second term can be omitted. 
The term of $k=l-1$ is only one remaining of the terms with $k \in \{ 0, \cdots , l-1 \}$. 
Hence we can estimate the second term as 
\begin{eqnarray}
    & & \frac 1N (\prod_{k'=0}^{l-2} U_{t-k'} )
        (1-\prod_{k'=0}^{l-1}U_{t-k'})^{-1}
        \sum_{\nu \geq 2}^{\alpha N/l} \sum_{\mu =1}^l
        \xi_i^{\nu \mu +1} \xi_j^{\nu \mu -l+1} x_j^{t-l+1(\nu \mu -l+1)}
        \nonumber \\
    &=& \alpha (\prod_{k'=0}^{l-2} U_{t-k'} )
        (1-\prod_{k'=0}^{l-1}U_{t-k'})^{-1} x_i^{t+1} .
    \label{eq:hi-t2-2nd}
\end{eqnarray}
We assume the third term is normally distributed with mean 0 and variance $\sigma_t^2$ because of the independence of $x_i^{t (\nu \mu )}$ and $\xi_i^{\nu \mu}$. 
The variance $\sigma_t^2$ of the cross talk noise is estimated as 
\begin{equation}
    \sigma_t^2
      = \alpha (1-\prod_{k'=0}^{l-1}U_{t-k'})^{-2}
        \left[
        q^t
      + \sum_{k=1}^{l-1}
        (\prod_{k'=0}^{k-1} U_{t-k'})^2
        q^{t-k}
        \right] ,
\end{equation}
where
\begin{equation}
    q^{t-k} = \frac 1N\sum_{j \ne i}^N (x_j^{t-k})^2 .
\end{equation}
The local field $h_i$ is obtained by setting $z_i \sim N(0,1)$ in (\ref{eq:hi-t2}), 
\begin{equation}
    h_i = \sum_{\mu =1}^s \xi_i^{\mu +1} m_\mu^t
        + \alpha (\prod_{k'=0}^{l-2} U_{t-k'} )
          (1-\prod_{k'=0}^{l-1}U_{t-k'})^{-1} x_i^{t+1}
        + \sigma_t z_i .
    \label{eq:hi-t4}
\end{equation}
The self-averaging property is assumed. 
Replacing $x_i \to Y$ and $\xi_i^\mu \to \xi^\mu$, we obtain the macroscopic equations as follows, 
\begin {eqnarray}
    Y^{t+1}(\xi^1, \cdots ,\xi^l;z)
      &=& F(\sum_{\mu =1}^l \xi^{\mu +1} m_\mu^t
                + \Gamma Y^{t+1}(\xi^1, \cdots ,\xi^l;z) + \sigma_t z ) \\
    m_\mu^{t+1} &=& \int^\infty_{-\infty}Dz\;
                \ll \xi^\mu Y^{t+1}(\xi^1, \cdots ,\xi^l;z) \gg  \\
    q^{t+1} &=& \int^\infty_{-\infty}Dz\; \ll
                {Y^{t+1}(\xi^1, \cdots ,\xi^l;z)}^2 \gg \\
    U^{t+1} &=& \frac 1\sigma_t \int^\infty_{-\infty}Dz \;
                z \ll Y^{t+1}(\xi^1, \cdots ,\xi^l;z) \gg \\
    \sigma_{t+1}^2 &=&
                \alpha (1-\prod_{k'=0}^{l-1}U_{t-k'})^{-2}
                \left[ q^t + \sum_{k=1}^{l-1}
                (\prod_{k'=0}^{k-1} U_{t-k'}) q^{t-k} \right] \\
    \Gamma  &=& \alpha (\prod_{k'=0}^{l-2} U_{t-k'} )
                (1-\prod_{k'=0}^{l-1}U_{t-k'})^{-1}
\end {eqnarray}
Now let us consider the case that the state of memory retrieval is periodic. 
We can set 
\begin{equation}
    m_\nu^t=m \delta_{\nu t}, \; \;
    q_t=q, \; \; U_t=U, \; \;
    \sigma_t=\sigma, \; \;
    (t=1, \cdots ,l) ,
\end{equation}
where $\delta_{\nu t}$ denotes Kronecker's delta. 
Finally, we obtain the macroscopic equations as follows: 
\begin {eqnarray}
    Y(\xi^{t+1};z)
      &=& F(\xi^{t+1} m
            + \Gamma Y(\xi^{t+1};z) + \sigma z )
          \label{eq:opeq-limit1} \\
    m &=& \int^\infty_{-\infty}Dz\;
          \ll \xi^{t+1} Y(\xi^{t+1};z) \gg \\
    q &=& \int^\infty_{-\infty}Dz\;
          \ll {Y(\xi^{t+1};z)}^2 \gg \\
    U &=& \frac 1\sigma \int^\infty_{-\infty}Dz \;
          z \ll Y(\xi^{t+1};z) \gg \\
    \sigma^2 &=& \alpha \frac {1-U^{2l}} {(1-U^l)^2(1-U^2)} q \\
    \Gamma &=& \frac {\alpha U^{l-1}} {1-U^l} .
          \label{eq:opeq-limit2}
\end {eqnarray}
Setting $F( \cdot )= \sgn ( \cdot )$ and using Maxwell rule \cite{shiino1992}, we obtain (\ref{eq:sna_m})-(\ref{eq:sna_rho}) as follows: 
\begin{eqnarray}
m &=& {\rm erf} \biggl( \frac m{\sqrt{2\alpha \rho}} \biggr), \\
U &=& \sqrt{\frac 2{\pi\alpha \rho}}e^{-\frac {m^2}{2\alpha \rho}}, \\
\rho &=& \frac{1-U^{2l}}{(1-U^2)(1-U^l)^2}, 
\end{eqnarray}
where $\sigma^2=\alpha \rho q$, $q=1$ and $\sgn (\cdot)$ denotes the sign function ($\sgn (x)=1$ for $x \ge 0$, $-1$ for $x<0$). 
Thus, we find that these stationary state equations of the order parameters given by the signal-to-noise analysis (SCSNA) 
are equivalent to those of the path-integral analysis (\ref{eq:sna_m})-(\ref{eq:sna_rho}).

\end{document}